\documentclass[useamsfonts]{pasj00}
\usepackage{mathrsfs}
\newlength{\mylength}

\begin{document}
\SetRunningHead{T.~Musil \& V.~Karas}{ADAF in the Kerr metric: 
 propagation of a wave pattern}
\Received{2002/06/29}
\Accepted{2002/07/22}

\title{Advection-dominated accretion flow in the Kerr metric:\\
 Propagation of a wave pattern}

\author{Tom\'a\v{s} \textsc{Musil} and Vladim\'{\i}r \textsc{Karas}}
\affil{Astronomical~Institute, Charles~University,
 V~Hole\v{s}ovi\v{c}k\'ach~2, CZ-180\,00~Prague,
 Czech~Republic}

\maketitle
\begin{abstract}
Time evolution of an advection-dominated accretion flow is explored
in terms of relativistic fluid equations. An axially symmetric, 
vertically averaged mean flow is constructed and then perturbed. 
An axisymmetric pattern is followed as it propagates in the form of 
a wave towards the horizon of a rotating (Kerr) black hole. 
Several assumptions are relaxed in comparison with previous works
(Manmoto et al.\ 1996). The wave reflection and steepening are 
examined and the influence of black hole angular momentum is 
discussed.
\end{abstract}

\section{Introduction}
In this paper we report on computations in which relativistic
hydrodynamics has been employed to describe propagation of a wave
pattern in accreted fluid. An axisymmetric perturbation is imposed on a
background model of a transonic, advection-dominated flow. The fluid has
negligible self-gravity; it moves in spacetime of a Kerr black hole that
is described by parameters $M$ and $a$.\footnote{Hereafter, equations
will be written in geometrized units and scaled with the central mass
$M$. Hence, dimensionless angular momentum acquires values
$0{\leq}a\leq1$. Corresponding quantities in physical units can be
obtained by straightforward conversions (Misner et al.\ 1993, p.~36).
Gravitational radius, $R_{\rm{g}}=GM/c^2$, will be adopted as a
convenient unit of length.} While our approach is essentially
one-dimensional, it is able to treat non-linear regime of the growing
perturbation. The model can provide some insight into phenomena expected
to occur in astrophysically realistic models.

We build our computations on the results from several earlier papers.
Particularly relevant have been computations of Manmoto et al.\
(1996; cited as P1 hereafter) who explored wave propagation
and reflection in a transonic flow (see also Kato et al.\
1998, chapt.~11). These authors found clear evidence for
reflecting waves near the horizon. The perturbation then steepens into a
shock wave. Observational motivation for these studies comes from the
evidence for X-ray fluctuations, origin of which is often linked with
disturbances in the innermost regions of accretion flows. In P1, the
background accretion flow was constructed in the pseudo-Newtonian
framework, while we employ general relativistic description and relax 
several other restrictions. 

The main aim of this note is to show the results of our computations,
where black hole rotation stands as one of the parameters. We
refer to other, more extended papers for a thorough discussion of
various connections and the place that these computations may have in a
general scheme of the black-hole accretion. For instance, a lot of interest
has been focused on the possibility of limit-cycle oscillations in accretion
disks (Honma et al.\ 1991). Indeed, non-linear time-dependent calculations of vertically
integrated transonic accretion disks exhibit limit-cycle
behaviour associated with thermal instability (Szuszkiewicz \& Miller 1998). 
Recently,
Bate et al.\ (2002) analyzed axisymmetric waves with a two-dimensional
hydrodynamical scheme.

This paper is organized as follows. In Sec.~\ref{sec:background} the 
background stationary flow is constructed. In Sec.~\ref{sec:perturbed} a
perturbation is imposed on the background flow and its time evolution
is examined. Finally, the properties of the temporal behaviour are summarized 
in Sec.~\ref{sec:results}.

\section{Equations and the method of solution}
We assume Kerr black-hole spacetime\footnote{We use Boyer-Lindquist
coordinates ($t,r,\theta,\phi$; 
Misner et al.\ 1973) with usual notation;
$\Sigma=r^2+a^2\cos^2\theta$, $\Delta=r^2-2r+a^2$, 
$A=(r^2+a^2)^2-{\Delta}a^2\sin^2\theta$, ${\mathscr A}=1+a^2/r^2 + 2a^2/r^3$,
${\mathscr C}=1-3/r+2a/r^{3/2}$, ${\mathscr D}=1-2/r+a^2/r^2$;
$g=\Sigma^2\sin^2\theta$ is determinant of the metric.
Horizon is at $r_+=1+\sqrt{1-a^2}$, $\theta=\pi/2$ is the flow
mid-plane.} in which hydrodynamic
equations are solved. For the background flow, we adopt 
a model described by Gammie \& Popham (1998) and Popham \& Gammie (1998).
Steady-state equations are summarized in
subsection \ref{sec:background} for convenience, 
while our main attention will be focused
on time evolution of a perturbation discussed later on
in the text. 

\subsection{A steady background accretion flow}
\label{sec:background}
The steady-state equations can be derived from:\vspace{2pt}

{\parindent0pt\parskip2pt

\parbox[t]{15pt}{(i)}~\parbox[t]{0.89\mylength}{Particle conservation 
$\left(\rho u^\mu\right)_{\!;\mu}=0$ which, thanks to stationarity,
axial symmetry and vertical averaging can be expressed in terms of
accretion rate: $\dot{M}=-4{\pi}H_{\!\theta}r^2\rho\,u^r$.}\hfill

\parbox[t]{15pt}{(ii)}~\parbox[t]{0.89\mylength}{Energy 
conservation $u_\mu{T^{\mu\nu}}_{;\nu}=0$, which can be written in the form
$u^r u_{,r}-(u+p)\rho^{-1}u^r\rho_{,r}=\Phi-\Lambda{\equiv}f\Phi$, where
$\Phi=-t^{\mu\nu}\sigma_{\mu\nu}$; $f$
represents the ratio of heating and cooling processes, respectively.}\hfill

\parbox[t]{15pt}{(iii)}~\parbox[t]{0.89\mylength}{Radial component of the 
momentum conservation law, 
$h_{r\mu}{T^{\mu\nu}}_{;\nu}=0$.}\hfill

\parbox[t]{15pt}{(iv)}~\parbox[t]{0.89\mylength}{Angular momentum conservation, 
$T^{\mu\nu}\xi_{\mu;\nu}=0$.}

In analogy with (Abramowicz et al.\ 1997), 
we defined the aspect ratio of the flow,
$H_{\!\theta}=H/r$ (where $H$ is geometrical semi-thickness);
$h_{\alpha\beta}$ denotes the projection tensor, $\xi_\mu$ is the axial 
Killing vector, and $\sigma_{\mu\nu}$ is the equilibrium shear stress 
tensor.}

Four equations correspond to conservation laws (i)--(iv):
\begin{eqnarray} \label{eq1}
4\pi r^2 \rho \, H_\theta V\;\sqrt{{\mathscr D}}\;\gamma_r &=& - \dot{M},\\
V\;\sqrt{{\mathscr D}}\;\gamma_r \left(\frac{\partial u}{\partial T}
\frac{{\rm{}d}T}{{\rm{}d}r}-\frac{p}{\rho}\frac{{\rm{}d}\rho}{{\rm{}d}r}\right)
 &=& f \Phi, \\
-\frac{1}{\rho\eta}\frac{{\rm{}d}p}{{\rm{}d}r} - 
{\gamma_\phi^2}\,r^{-2}{\mathscr A}{\mathscr D}^{-1}
{\mathscr Z} &=& \frac{V}{1-V^2}\frac{{\rm{}d}V}{{\rm{}d}r},\label{eq3}\\
\dot{M}l\eta - 4\pi H_\theta\,r^2\, t^r_{\,\phi} &=& \dot{M}j.
\label{eq4}
\end{eqnarray}
The unknowns are: mid-plane density and temperature, $\rho$ and $T$, 
and two components of the fluid four-velocity $u_\mu$.\footnote{One of 
the two components can be chosen in the azimuthal direction, 
$l=u_\phi$, the latter is conveniently
expressed as the radial speed $V$ of the fluid local rest frame ({\sf{}LRF}) 
with respect to the local frame co-rotating with the fluid 
at constant radius ({\sf{}CRF}). In eq.~(\ref{eq3}), $\gamma_\phi$
is the Lorentz factor corresponding to azimuthal boost between {\sf{}CRF}
and the locally non-rotating frame ({\sf{}LNRF}) and similarly
$\gamma_r=(1-V^2)^{-1/2}$.} 
In eq.~(\ref{eq3}), we denoted ${\mathscr Z}=1-2a\Omega-\left(r^3-a^2\right)\Omega^2$,
$\Omega=u^\phi/u^t$. In eq.~(\ref{eq4}), $j$ is the eigenvalue of the
problem and $t^r_{\,\phi}$ is the viscous tensor component. See
Gammie \& Popham (1998) and the Appendix for details of the derivation. 

Equations (\ref{eq1})--(\ref{eq4}) were solved numerically and the obtained
results agree with Gammie \& Popham (1998). 
The steady solution will be used as the
initial state of the time-dependent flow integration in the next subsection.

\begin{figure}[tb]
\begin{center}
\FigureFile(78mm,78mm){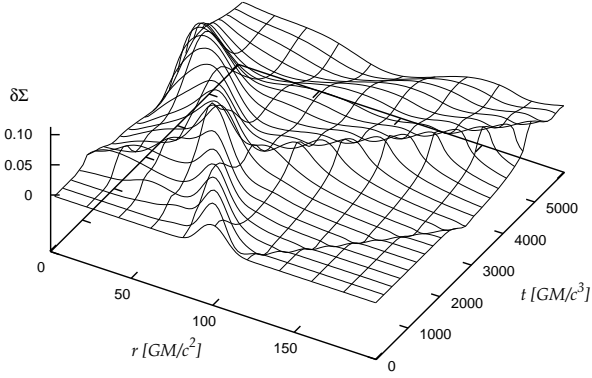}
\end{center}
\caption{Time evolution of surface density perturbation in our model 
($\delta\Sigma$ in units of g/cm$^2$). 
Schwarzschild metric was assumed 
with $\alpha=0.1$, $f=1$.}\label{fig1}
\end{figure}

\subsection{Time evolution of a perturbed flow}
\label{sec:perturbed}
Equations describing evolution of the flow are equivalent to those
listed in items (i)--(iv) in the previous subsection, except that we do
not neglect terms containing time derivatives. 

The initial imposed perturbation is adopted analogical as in ref.\ P1
(sec.~2), so that comparisons are easier. Namely, for the density
perturbation we assume
$\delta\rho/\rho=\kappa_0\exp\left[-(r-r_0)^2/(\lambda_0/2)^2\right]$,
where $\kappa_0<1$ is the starting perturbation amplitude, $r_0$ is its
location and $\lambda_0$ is the characteristic length. Notice that
density perturbation can be related to perturbations of other quantities
-- $T$, $l$, and $V$ (Kato et al.\ 1996). An explicit form of the equations is
given in the Appendix. We solved the set (\ref{eq5})--(\ref{eq8}) using
the Lax-Wendroff scheme (Press et al.\ 1992) with the free ingoing boundary
condition imposed close above $r=r_+$ (below the sound point). Far from
the black hole, the asymptotic form of the flow goes over to the
stationary solution, eqs.~(\ref{eq1})--(\ref{eq4}).

\section{Results: Waves propagation and reflection}
\label{sec:results}
First, we reproduce (in the general relativistic framework) the 
basic behaviour of the flow perturbation. Namely, we can observe
the initial propagation of the wave towards horizon, its reflection
in the region of several $R_{\rm{}g}$ (near the sonic point) and then
formation of a shock.

\begin{figure*}[tb]
\hfill
\FigureFile(76mm,76mm){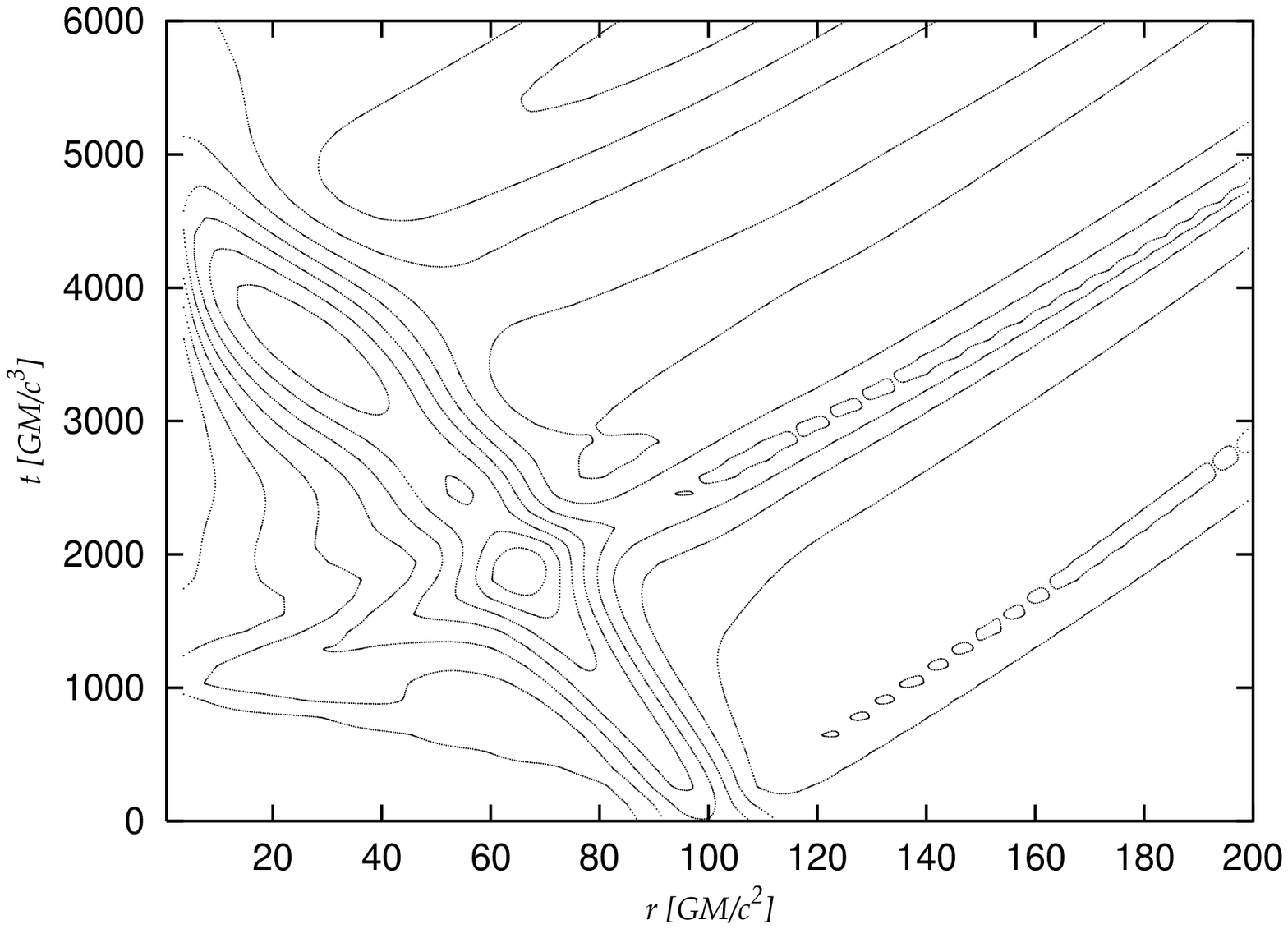}
\hfill
\FigureFile(76mm,76mm){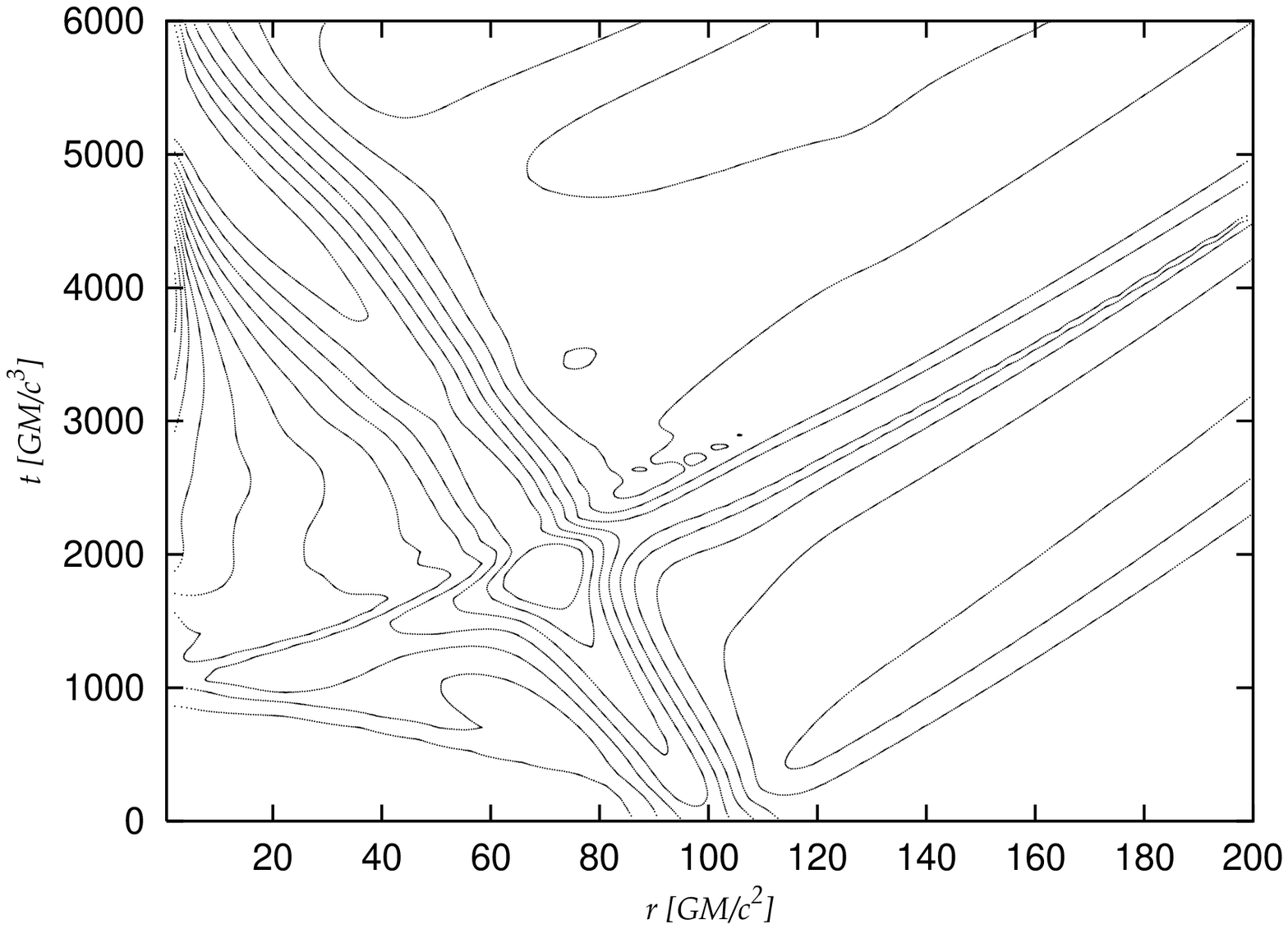}
\hfill
\caption{Surface density perturbation for $a=0$ (left)
and $a=1$ (right). The perturbation is imposed at $r=100$ (time $t=0$)
and it proceeds towards the black hole. The first outgoing wave is
reflected at about $t=1500$ where contour-lines indicate the
maximum perturbation. Further waves appear at
later times. Contours are plotted with equidistant spacing.
Parameters $\alpha$ and $f$ as in Fig.~\ref{fig1}.}\label{fig2}
\end{figure*}

In Figure~\ref{fig1}, perturbation $\delta\Sigma(r,t)$ of the surface
density $\Sigma=2rH_\theta \rho$ is plotted. We set $M=10M_{\odot}$ and
$\dot{M}=0.16 L_{\rm{}E}/c^2=2.24\times10^{17}\mbox{g s}^{-1}$ for 
definiteness. The integration domain was $300R_{\rm{}g} \times 10^4R_{\rm{}g}/c$
(somewhat smaller extent is captured in the graph; $R_{\rm{}g}/c$
corresponds to $\sim10^{-4}$s for $M=10M_{\odot}$). This figure can be
compared with Fig.~1 in P1 (after a trivial scaling of units). Our
integration time is longer, allowing us to observe subsequent outgoing
waves appearing at $t\gtrsim3\,000$ ($\approx 10$ viscous periods at the
outer edge of the integration grid). For definiteness we adopted $a=0$
here.

In Figure~\ref{fig2}, $\delta\Sigma(r,t)$ is shown for two limiting
values of angular momentum, $a=0$ and $a=1$, respectively. Contour-lines
show a crest of the initial perturbation which proceeds with the
accretion flow. At small radii, pattern speed is somewhat influenced by
black hole angular momentum. Several reflected waves appear at later
stages until the simulation settles back to the initial state. Given
the initial perturbation and keeping $\alpha$ and $f$ fixed, the form and
the amplitude of waves are slightly sensitive to the black hole angular 
momentum -- the effect that is difficult to treat within the 
pseudo-Newtonian approach. However, the general scheme is in good 
agreement with the pseudo-Newtonian treatment of paper P1.

\section{Conclusions}
We examined temporal behaviour of a simple relativistic model of the
advection-dominated flow. To this aim, we observed the evolution of
disturbances originating in outer parts of the disk. Reflection of
the waves was discussed previously in various contexts. We
developed a robust code which is able to deal with the problem across the 
large range of disk parameters without introducing numerical viscosity.
The scheme appears useful as a test bed for more complicated
investigations with unstable behaviour, treated in more dimensions
(work in progress).

In this note we described a stable system which after a transition period
eventually relaxes to the initial steady state. Our scheme, thanks to
its rather general formulation, exhibits more complex behaviour of
reflected waves than the pseudo-Newtonian model described in P1. The
main advantages are general relativistic treatment of hydrodynamic
equations including finite propagation speed of viscous effects. We are
also able to follow the evolution for longer intervals, till the 
shock escapes from the observed region and the model settles back 
to its initial equilibrium state.

VK acknowledges helpful discussion with Professor Marek Abramowicz. Support
from the grants GACR 205/00/1685 and 202/02/0735 is also acknowledged.

\appendix
\section*{~}
The only non-vanishing component of the viscous tensor (in {\sf{}LRF}),
$S=t_{(r)(\phi)}$, satisfies equation 
$u^{\mu}S_{;\mu}=-\tau_{\rm{}r}^{-1}(S-S_0)$ 
where $\tau_{\rm{}r}$ is the fluid relaxation time. 
An equilibrium value $S_0$ is given in terms of $\alpha$ parameter, 
speed of sound $c_{\rm{}s}$, enthalpy $\eta=(\rho+u+p)/\rho$, 
and the shear tensor component $\sigma\equiv\sigma_{(r)(\phi)}$:
$S_0=-2\rho\eta\nu\sigma$, where $\nu=\alpha{c_{\rm{}s}}rH_\theta$. 
Energy conservation can be further adapted using equation of 
state, $p={\rho}T$, where $\rho$ is the rest mass density and 
$T$ is dimensionless temperature 
($T={k}T_{\rm{}k}/\overline{m}c^2$ where $T_{\rm{}k}$ is kinetic 
temperature and $\overline{m}$ is the mean molecular mass). 

We summarize time-dependent equations of
sec.~\ref{sec:perturbed}:
\begin{eqnarray} \label{eq5}
\frac{\partial}{\partial t}\left(H_\theta \rho \, u^t\right) +
\frac{2}{r} H_\theta \rho \, u^r + \frac{\partial}{\partial
r}\left(H_\theta \rho \, u^r\right) &=& 0,  
 \\
\rho \left(g+T\frac{{\rm{}d}g}{{\rm{}d}T}\right) \left(u^t\,\frac{\partial T}
{\partial t} + u^r\,\frac{\partial T}{\partial r}\right) - T u^t\,
\frac{\partial \rho}{\partial t} && \nonumber \\
- T u^r\,\frac{\partial\rho}{\partial r} 
 +2\,f S\,\sigma_{(r)(\phi)} &=& 0 , 
 \\
\frac{\gamma_r V}{\sqrt{\mathscr D}}\,u^t\,\frac{{\rm{}d}p}{{\rm{}d}t}  +
\rho\,\eta\,u^t\,\frac{\gamma_r^3}{\sqrt{\mathscr D}}
\frac{{\rm{}d}V}{{\rm{}d}t} +
\gamma_r^2 \frac{{\rm{}d}p}{{\rm{}d}r}  && \nonumber \\ 
+ \rho\,\eta\,\gamma_r^4 V\,\frac{{\rm{}d}V}{{\rm{}d}r}
+ \rho\,\eta\,\left(\frac{u^t}{r}\right)^2 {\mathscr Z}&=& 0, 
 \\
H_\theta \rho\, u^t\,\frac{\partial l \eta}{\partial t}  +
\frac{\partial}{\partial t} \left( H_\theta\, t^t_{\,\phi}
\right) + \frac{2}{r} H_\theta\, t^r_{\,\phi} &&  \nonumber \\
+ H_\theta \rho\, u^r\,\frac{\partial l \eta}{\partial r}\, 
+ \,\frac{\partial}{\partial r} \left( H_\theta\, t^r_{\,\phi}
\right) & = & 0, 
 \\
u^t\,\frac{\partial S}{\partial t} + 2 \rho \, \eta \, \alpha\,
c_{\rm{}s}^2\, \left[\frac{l \gamma_r^3}{r\,\sqrt{{\mathscr D}}}
\left({\textstyle{\frac{1}{2}}}-V^2\right) \frac{\partial V}{\partial t} 
 \right.\nonumber \\
 + \left.\frac{V \gamma_r}{2r\,\sqrt{{\mathscr D}}} \frac{\partial
l}{\partial t} \right] + u^r\,\frac{\partial
S}{\partial r} + \frac{c_{\rm{}s} S}{r H_\theta} + 2 \rho \, \eta \,
\alpha\, c_{\rm{}s}^2\, \sigma & = & 0.
\label{eq8}
\end{eqnarray}
Here, the unknown functions ($T$, $l$, $V$, $\rho$, $S$) all depend on
$t$ and $r$ (variables $l$ and $V$ were introduced in footnote~3). 
Geometrical semi-thickness of the flow is given by
$\rho\eta\left[l^2-a^2\left(u_t^2-1\right)\right]H_\theta^2=pr^2$. Time
dependence of $H_\theta$ can be expressed via $T$, $l$ and $V$. Finally,
for the dissipation function one finds $\Phi=-2S\sigma_{(r)(\phi)}$ with
$\sigma_{(r)(\phi)}$ being the {\sf{}LRF} time-dependent shear component,
which is of the same form as in steady state ($\sigma_0$) plus correction 
terms:
\begin{equation} \label{sigmaerfi22}
\sigma_{(r)(\phi)} = \sigma_0 + \frac{l \gamma_r^3}{r\,
\sqrt{{\mathscr D}}} \left(\textstyle{\frac{1}{2}}-V^2\right)
\frac{\partial V}{\partial t} +
\frac{V \gamma_r}{2r\,\sqrt{{\mathscr D}}} 
\frac{\partial l}{\partial t}.
\end{equation}

\noindent
{\bf References}

\noindent
   Abramowicz M.\,A., Lanza A., Percival M.\,J.\ 1997, ApJ 479, 179 \\
   Bate M.\,R., Ogilvie G.\,I., Lubow S.\,H., Pringle J.\,E.\
   2002, MNRAS 332, 575 \\
   Gammie C.\,F., Popham R.\ 1998, ApJ 498, 313 \\
   Honma F., Kato S., Matsumoto R.\ 1991, PASJ 43, 147 \\
   Kato S., Abramowicz M.\,A., Xinming C.\ 1996, PASJ 48, 67 \\
   Kato S., Fukue J., Mineshige S.\ 1998, Black-Hole Accretion 
   Disks (Kyoto University Press, Kyoto) \\
   Manmoto T., Takeuchi M., Mineshige S., Matsumoto R., Negoro H.\
   1996, ApJ 464, L135 (paper P1) \\
   Misner C.\,W., Thorne K.\,S., Wheeler J.\,A.\ 1973,
   Gravitation (Freeman, San Francisco) \\
   Popham R., Gammie C.\,F.\ 1998, ApJ 504, 419 \\
   Szuszkiewicz E., Miller J.\,C.\ 1998, MNRAS 298, 888 \\
   Press W.\,H., Teukolsky S.\,A., Vetterling W.\,T.,
   Flannery B.\,P.\ 1992, Numerical Recipes 
   (Cambridge University Press, Cambridge)

\end{document}